# Identifying 21st century STEM competencies using workplace data

Hyewon Jang [1]

[1] *John A. Paulson School of Engineering and Applied Sciences, Harvard University, Cambridge, Massachusetts 02138*

## Abstract

Gaps between Science, Technology, Engineering, and Mathematics (STEM) education and required workplace skills have been identified in industry, academia, and government. Educators acknowledge the need to reform STEM education to better prepare students for their future careers. We pursue this growing interest in the skills needed for STEM disciplines and ask whether frameworks for 21st century skills and engineering education cover all of important STEM competencies. In this study, we identify important STEM competencies and evaluate the relevance of current frameworks applied in education using the standardized job-specific database operated and maintained by the United States Department of Labor. Our analysis of the importance of 109 skills, types of knowledge and work activities, revealed 18 skills, seven categories of knowledge, and 27 work activities important for STEM workers. We investigate the perspectives of STEM and non-STEM job incumbents, comparing the importance of each skill, knowledge, and work activity for the two groups. We aimed to condense dimensions of the 52 key areas by categorizing them according to the Katz and Kahn (1978) framework and testing for inter-rater reliability. Our findings show frameworks for 21st century skills and engineering education do not encompass all important STEM competencies. Implications for STEM education programs are discussed, including how they can bridge gaps between education and important workplace competencies.

## I.    INTRODUCTION

Jana graduated from university with a Science, Technology, Engineering, and Mathematics (STEM) degree. In her interview with a leading engineering firm, Jana's interviewers asked about her experience relevant to the skills for the job. Had her education afforded her relevant experience? If STEM education programs are to meet the demand for skills in society, they must comprehend the nature of skills required in the workplace to better prepare graduates.

However, gaps between STEM education and required skills in the workplace have been reported and discussed in industry, government, and academia (Hung-Lian et al. 2000; Lee and Fang 2009; Lu et al. 1998; Meier et al. 2000). Written and oral communication, project management, teamwork, problem solving, critical thinking, and interpersonal skills are frequently reported as lacking in STEM graduates (Hung-Lian et al. 2000; Lu et al. 1998; Radermacher and Walia 2013). More concerning is that frameworks designed for education have not been empirically evaluated with respect to workplace relevance. The engineering education criteria, developed by the Accreditation Board for Engineering and Technology (ABET) and widely applied in universities in the United States of America (Felder and Brent 2003; Prados et al. 2005), have been described as a failure from the perspective of practicing engineers (Jonassen et al. 2006; Petroski 1996).

In light of the importance of bridging gaps between STEM education and skills, we investigated important STEM competencies for the 21st century using database collected directly from employees managed by the United States (U.S.) Department of Labor. We posed three research questions: 1) What are the most important competencies for STEM job incumbents in the 21st century? 2) Compared with non-STEM disciplines, what competencies are significantly more important for STEM disciplines? 3) Do frameworks relevant to 21st Century Skills and engineering education completely identify important STEM competencies?

STEM educational policies grew out of concerns about the STEM workforce (PCAST 2010; NRC 2008), including the predicted shortage of workers and lack of skills in this group. With respect to the shortage of STEM workers, researchers have explored factors that affect the pipeline. For example, early life experiences, classroom experiences, college course enrollment, choice of academic major, college persistence, gender and race, and career aspirations were found to be associated with obtaining STEM degrees (Astin and Astin 1992; Bonous-Hammarth 2000; David T. Burkam 2003; Maltese and Tai 2011; Maple and Stage 1991; Tai et al. 2006; Trusty 2002). The number of pedagogical methods used, such as hands-on activities, coverage of topics more relevant to students, and greater use of cooperative learning strategies, was reported to improve the attitudes of students to STEM fields (Myers and Fouts 1992; Piburn and Baker 1993). Building on these findings, educational efforts to construct curricula with interdisciplinary collaboration, active learning, and learner-centered instruction have been stressed as core strategies for change.

However, those efforts have not shown substantial success when considering the significant resources invested in research and development to improve STEM education. Research has emphasized that a fundamental shift in the entire education paradigm from teacher-centered to student-centered is necessary (Henderson 2008). Not surprisingly, education reform efforts to improve STEM classroom experiences have typically addressed teaching, student learning, and student learning productivity (Fairweather 2008). To leverage change in instructional practices, strategies to change the beliefs of the individuals should be involved (Henderson et al. 2011). Interestingly, engineering educators in Massachusetts Institute of Technology (MIT) explored two central questions for building engineering education standards (Crawley et al. 2007): "what is the full set of knowledge, skills, and attitudes that engineering students should possess as they leave the university, and at what level of proficiency?" (p. 10) and "how can we do better at ensuring that students learn these skills?" (p. 10) They believed that education was to equip students to become successful STEM job incumbents.

In this study we explore important skills, knowledge, and work activities using the standardized occupation information database managed by the Department of Labor. In addition, we compare it with frameworks of 21st century skills and the engineering criteria (2015~2015) (ABET 2015), to discuss the need to improve current frameworks and develop a more complete set of important STEM competencies. These frameworks have been used to address skills needed for life and career success, and consequently they have impacted on STEM education. A framework of 21st century skills was created in a workshop held by the National Research Council (2008), with the aim of preparing students for future workplaces. Technological developments mean that computers have increasingly replaced humans in performing routine tasks (Koenig 2011). Workers are required to have cognitive and social skills relevant to non-routine problem solving and complex communication. Experts suggested five 21st century skills recognized as increasingly important in the labor market: adaptability, complex communication skills, non-routine problem-solving skills, self-management/development, and systems thinking (Koenig 2011). In addition to the National Research Council, several other groups, such as Partnerships for 21st Century Skills, have developed frameworks identifying the most essential skills for the future workforce. This has resulted in educators involving the Department of Education to support changing the standard for educating students and designing assessments to measure these standards (Koenig 2011). However, frameworks of 21st century skills are still seldom empirically examined from a STEM job incumbent's perspective.

Engineering criteria (ABET 2015) have led to changes in curricula and instructional methods for engineering education (Prados et al. 2005). Engineering education has attempted to ensure that engineering graduates are well prepared for professional practice, using educational quality control over the past 80 years. Employers and educational leaders in this field contributed to education criteria to ensure the effective preparation of engineers. For example, employers reported weaknesses of engineering graduates, such as difficulties working in a team. Educational leaders developed models to retain the strengths of engineering science, but to remedy these specific reported weaknesses (Prados et al. 2005). Since 2000, ABET have implemented outcomes-based criteria to characterize the skills, knowledge, and attitudes engineering graduates should have (Table 6). Engineering criteria (ABET 2015) have been used for accreditation to evaluate whether an engineering program achieves a minimum level of competence (Prados et al. 2005). Interestingly, engineering criteria have not been empirically evaluated with respect to STEM workplace relevance. Undoubtedly, professionals who created the framework of 21st century skills and engineering criteria are knowledgeable about required skills and practices. However, empirical investigation of these frameworks using standardized occupation information is still needed.

This paper is structured as follows: Theoretical framework, methods, and data are described in section II. In section III, we present results, showing important skills, knowledge, and work activities for STEM disciplines. We categorize important skills, knowledge, and work activities into a framework and compare them with those detailed in the literature. We summarize and generalize our findings and discuss implications for STEM education in section IV. Overall, these results may be used to inform STEM education programs about how to better prepare students for the workplace, thereby helping bridge the gap between STEM education and the required skills for STEM job incumbents.

## II. METHODS

### A. Theoretical framework of job information

We used the O*NET database: a standardized, job-specific database operated and maintained by the U.S. Department of Labor. To investigate 21st century STEM competencies, it is important to use recent data collected directly from employees or job analysts. The database is based on job analysis and seeks to form the foundation of a wide range of human resource management. Applied psychologists analyzed the work activities that characterize jobs (Peterson et al. 1999) with a general objective "to define the fewest independent ability categories which might be most useful and meaningful in describing performance in the widest variety of tasks" (p. 352) (Fleishman 1967). Many methods have been used for developing job descriptions. These procedures have changed over time, however the tools used in job analysis, such as observations by job analysts, surveys for job incumbents, and interviews at one or more sites, have stayed the same (Peterson et al. 1999).

Job analysis is based on a task analysis. A task can be defined as "any set of activities, occurring at the same time, sharing some common purpose that is recognized by the tasks performer" (p. 11) (Miller 1967). Tasks are classified by the human capacities that are necessary to perform them effectively. Job analysis is based on taxonomies of human performance considering four approaches: 1) the *Behavior Description Approach* (categories of tasks are formulated based on observations and descriptions of what operators actually do while performing tasks), 2) the *Behavior Requirements Approach* (behaviors should be carried out, or are required, to achieve criterion levels of performance), 3) the *Ability Requirements Approach* (tasks are to be described, contrasted, and compared in terms of abilities that a given task requires of the individual performer), and 4) the *Task Characteristics Approach* (a task is a set of conditions that elicits performance) (Fleishman and Quaintance 1984).

For completing tasks in any job, skills are required at the level of proficiency for a specific task or group of tasks. Skills represent a performance domain involving the acquisition and application of certain types of knowledge. Initial research into the nature of skilled performance defined skills operationally: in terms of gains in performance with practice on a certain task (Peterson et al. 1999). The primary objective of these early studies was to identify the variables contributing to more rapid acquisition of skilled performance (Fleishman 1967, 1972, 1982; Fleishman and Hempel Jr 1955). After these studies, techniques such as protocol analysis and comparison of expert-novice differences were used to identify the key characteristics of skilled performance (Chi and Glaser 1985). Skilled performance appears to involve acquisition of a set of strategies, procedures, and processes associated with expertise (Ackerman 1988; Fleishman and Hempel Jr 1955; Ward et al. 1990).

In this study, we focus on skills and competencies rather than abilities. Skills and competencies are more situational, more dependent on learning, and represent the product of training in particular tasks or individual attributes related to quality of work performance. Abilities are defined as "…relatively enduring attributes of an individual's capability for performing a particular range of different tasks" (p. 175) (Peterson et al. 1999). Skills or competencies can be measured by the quality of relevant jobs at work. The improvement of a skill is partially related to an individual's possession of relevant underlying abilities (Ackerman 1988; Fleishman 1966, 1967, 1972).

***Development of database of job information***

The O*NET database has been developed through the historical, interdisciplinary efforts of business, government, and scholars of applied psychology. *The Dictionary of Occupational Titles (DOT)* project began in the 1930s to help the new public employment system during the economic depression and to improve links between skill supply and demand. Data were collected to identify skills, knowledge, abilities, and traits that workers needed, as well as the education and training requirements, machines, tools, equipment and materials used and products produced (Peterson et al. 2001). Following publication of the first DOT, updated versions had been published until 1991 to provide descriptive information for more than 12,000 occupations.

Although the DOT had been used for over 50 years, limitations eventually became clear. Information was outdated and overly job-specific, and the structure of the DOT did not easily facilitate comparisons across jobs (Markle et al. 2013). At the same time, social concerns about worker skills were growing in industry, government, and education. This concern was clear in a number of initiatives, including the Secretary's Commission on Achieving Necessary Skills (SCANS) (Labor 1992), which explored rapid changes in technology, global competition, and the emergence of new organizational structures. The Secretary of Labor acknowledged, "the real economic challenge facing the United States in the years ahead ... is to increase the potential value of what its citizens can add to the global economy, by enhancing their skills and capacities and by improving their means of linking those skills and capacities to the world market" (p. 4) (APDOP 1993). These challenges highlight the need for complex performance skills and ongoing skill development.

The Advisory Panel for the Dictionary of Occupational Titles (APDOT), consisting of prominent applied psychologists, recommended a new occupational system, noting that it should promote the effective education, training, counseling, and employment of the US workforce (APDOP 1993). The APDOT proposed specific recommendations for the new DOT, categorized according to the issues of purpose, database, data collection, dissemination and implementation. The new occupational information system was designed to address a wide range of queries, be maximally useful, and capable of providing accurate descriptions of work characteristics and worker attributes. In 1995, the US Department of Labor first used the term *Occupational Information Network* (O*NET) to describe its new occupational tool.

O*NET is a comprehensive database of workers and occupational characteristics based on the content model, a framework that defines key features of an occupation using a standardized, measurable set of variables called

'descriptors' (Peterson et al. 1999). The content model takes into account the nature of jobs, characteristics, and requirements. This model consists of six domains of worker-oriented and job-oriented descriptors: 1) *Worker characteristics, 2) Worker requirements, 3) Experience requirements, 4) Occupational requirements, 5) Workforce characteristics, and 6) Occupation-specific information*. The O*NET is yearly updated by surveying a range of workers and job analysts and is available to the public. Therefore, using the O*NET aligns with our goal to identify important 21st century competencies for future STEM job incumbents. Because ultimately they must adapt in constantly changing workplaces where they will be confronted with a host of complex new roles and asked to take responsibility for their own work and careers.

We focused on aspects of the *Worker requirements* and *Occupational requirements* domains because the purpose of our study is to identify 21st century competencies that are critical and trainable for STEM job incumbents. *Worker requirements* are defined as *work-related attributes acquired and/or developed through experience and education*. The *Worker requirements* domain includes three subdomains: *Skills, Knowledge*, and *Education*. The *Occupational requirements* domain refers to *a comprehensive set of variables or detailed elements that describe what various occupations require*. The *Occupational requirements* domain contains *Work activities (general, intermediate, and detailed), Organizational context*, and *Work context*. To choose relevant subdomains, we reviewed subdomains and descriptors with two questions: 1) Can this inform important, required skills or knowledge? 2) Can this help us to understand what STEM workers do in workplaces? After the review, we decided to focus on three subdomains: *Skills* and *Knowledge* in *Worker requirements* and *Work activities* in *Occupational requirements*. The rationale for this choice is that the subdomains of *Skills* and *Knowledge* inform us what capacities should be developed to facilitate performance across jobs. Similarly, the subdomain of *Work activities* informs us about work activities common across many occupations. In contrast, the subdomain of *Education* details only the required level of education, while subdomains of *Organizational* and *Work context* elucidate the characteristics, physical and social factors that influence work.

**B. Sample**

We defined STEM disciplines as *occupations that require education in science, technology, engineering, and mathematics (STEM) disciplines* based on the definition in the O*NET. STEM disciplines include chemistry, computer science, engineering, environmental science, geosciences, life sciences, mathematics, and physics/astronomy. We used the O*NET 18.1, which was released in March 2014. Of 923 occupations, 157 belonged to the STEM disciplines. The O*NET database has been used by other researchers (Burrus et al. 2013; Markle et al. 2013) and therefore this study can be compared with previous research.

The O*NET consisted of 1) a random sample of businesses expected to employ workers in the targeted occupations, and 2) a random sample of workers in those occupations (Peterson et al. 1999). Survey questions were organized into three questionnaires, each containing a different set of questions. The sampled job incumbents for each occupation were randomly assigned one of three questionnaires. All respondents were also asked to complete a task questionnaire and provide demographic information.

The sample consisted of a total of 50,527 individuals who were either employees chosen randomly or job analysts (Table 1). Using the definition of STEM disciplines, 9,950 individuals were identified as being involved in STEM fields: 1,240 job analysts participated in the survey of *Skills* (an average of eight people per occupation); 4,348 job incumbents took part in the survey of *Knowledge* (an average of 27 people per occupation), and 4,362 job incumbents completed the survey of *Work activities* (an average of 28 people for each occupation).

**Fig.1 (a), (b), (c)** Distributions of average importance of skills (n=35), knowledge (n=33), and work activities (n=41) in STEM disciplines. The dotted red line represents the median of importance of ratings for each domain, whereas the dotted blue line represents 2.95 points. **Fig. 1(d)** 52 descriptors categorized into the *High* group (red) and 57 descriptors grouped into the *Low* (blue) group

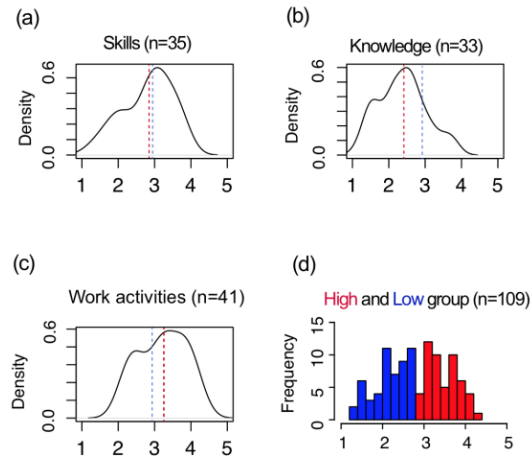

**C. Analysis**

Before investigating important STEM competencies, we divided the dataset into two groups: STEM and non-STEM disciplines using the occupation code and the definition of STEM disciplines. Three subdomains of *Skills, Knowledge, and Work Activities* were used to measure the importance of 35 skills, 33 types of knowledge, and 41 work activities. The following question was used: "How important is this <skill/knowledge/work activity> to the performance of your current job?" Ratings were made on a five-point scale (1=not important, 2=somewhat unimportant, 3=important, 4=very important, 5=extremely important). For example, with respect to *Knowledge* of *Mathematics*, job incumbents were asked, "How important is this knowledge of Mathematics to the performance of your current job?" To explore the distribution of each rating, we calculated the mean, median, and standard deviations for each subdomain (Skills: mean 2.81, median 2.91, SD 0.69; Knowledge: mean 2.19, median 2.09, SD 0.55; Work activities: mean 3.20, median 3.30, SD 0.66) and plotted the average importance of each rating for each domain (see Fig. 1(a)–(c)).

To choose important descriptors, 109 ratings were listed in order of average importance. The median of the ratings was 2.86. To categorize important descriptors into the *High (H)* group, we considered using either the median or the three points from the rating scale where three points signifies 'important' for the performance of the rater's current job. To select important descriptors rigorously, we used 2.95 points over the median (2.86), which is equal to 3 points after rounding off to two decimal places. After all ratings were divided into two groups, either *High* (above 2.95 points) or *Low* (below 2.95 points) (Fig. 1(d)), we used the Wilcoxon rank sum test (Hollander, 2013) for two-matched samples to affirm the difference between the *High* and *Low* groups, because average scores of the importance for the two groups were not normally distributed (Table 1). The null hypothesis is that there is no statistical difference between average importance of High (H) and Low (L) groups on matched descriptors; $H_0: H - L = 0$. The alternative hypothesis is that the average importance of the High group (H) is significantly higher than the average importance of the Low group (L) on matched descriptors; $H_0: H - L > 0$.

We also explored different perspectives of STEM and non-STEM job incumbents, because important competencies specific to the STEM disciplines had been documented. First we investigated whether or not there were differences in responses between STEM and non-STEM disciplines. Second, we reported the specific and important descriptors for STEM workers, and compared the importance of each descriptor and effect size between the two disciplines. In doing so, we classified relatively important skills, knowledge, and work activities for the STEM disciplines, which had an effect size over 0.3.

**D. Categorization**

Before categorization took place, we explored statistical methods using a merged dataset of three domains, such as principal component analysis (PCA). It was useful for condensing large data into a smaller dimensionality and interpreting the correlation among ratings. However, extracting the important STEM competencies from the results of PCA was complicated. This was first, because eight components were necessary to account for 70% of the total variance, and second, because the principal components were mismatched with results showing important skills, knowledge, and work activities. For example, the third component (SS Loading: 14.11; Proportion variance: 0.13) represented physical skills, explained by 13 items with loading over 0.5, such *as Installation, Handling and moving objects*. However, physical skills related to mechanical work were not evaluated as important by STEM workers. Indeed, the average importance of descriptors relevant to physical tasks, such as *Handling and moving objects* in work activities, scored under 2.5 points. PCA captured correlational relationships between 109 skills, knowledge, and work activities rather than identifying important competencies for STEM job incumbents.

To condense dimensions of important skills, knowledge, and work activities, we categorized them according to a framework developed by Katz and Kahn (1978), and verified the inter-rater reliability. To choose the best framework for categorization, we posed three questions: 1) Does the framework include all our findings? 2) Does the framework employ diverse domains of human performance? 3) Is the framework clear and simple for future job incumbents and educators to understand? We explored frameworks used to assess general human performance. The most relevant framework was that of Katz and Kahn (1978), which includes categories to represent skills and knowledge required in technology-based workplaces and relevant to working in an organization. This framework illustrates general human performance required in workplaces, such as solving problems, working with people, using technology, functioning in an organizational system, and working with resources. The framework explains work as "a process of transforming raw materials into useful products through the use of technology and labor (p. 50)" based on sociotechnical system theory (Peterson et al. 1999). It has five domains so that it is easily understood by future job incumbents and educators.

To categorize 52 important skills, knowledge, and work activities into five categories, we asked two raters, each with at least 5 years' experience in STEM education research, to review the operational definition of each descriptor and to choose the most applicable category. Each descriptor was categorized into a domain of the framework. For example, if a rater thought *Critical Thinking* would be most relevant to solving problems, then it would belong to the category of *Problem-solving Skills.* To ensure the reliability of the categorization, two raters performed the task separately, and then inter-rater reliability was calculated. In cases of disagreement, the two raters discussed the issue until a consensus was reached regarding the item's placement.

### D. Validity and reliability

To ensure validity and reliability, we reviewed the design and data collection procedures of the O*NET database. The methodology of the O*NET consists of empirical analysis and consensus of applied psychologists, educators, and incumbents. Analyses were completed using data from approximately 50,000 individuals in US workplaces who answered the O*NET questionnaire. To ensure reliability of categorization, as outlined in the preceding section of this paper, we reviewed frameworks and measured the inter-rater reliability. After categorization, we explored all descriptors in each category, reviewing educational literature, to find implications for STEM education.

### III.    RESULTS

In this section, first we report important skills, knowledge, and work activities for STEM job incumbents. The averages of importance for the descriptors in the *High* group were significantly higher ($p < 0.001$) than descriptors in the *Low* group. For example, the average importance level of skills in the *High* group was 3.35, whereas this figure was 2.23 in the *Low* group. These results show that descriptors in the *High* group can be regarded as important. To understand important competencies, here we refer to the operational definition (italicized) of each descriptor (italicized) in the O*NET.

**Table 1** *High* and *Low* skills, knowledge, and work activities. Skills, knowledge, and work activities over 2.95 points were categorized into the *High* group; others were grouped into the *Low* group; 18 skills, 7 types of knowledge, and 27 work activities were grouped into the *High* group and 17 skills, 26 types of knowledge, and 14 work activities were categorized into the *Low* group. To test differences statistically, we used the Wilcoxon rank sum test. In all cases, descriptors of High group score significantly higher than those of Low group.

| Group | *High* N (Mean, Median, SD) | *Low* N (Mean, Median, SD) | Wilcoxon *Z* statistic |
|---|---|---|---|
| Skills | 18 (3.35, 3.31, 0.29) | 17 (2.23, 2.15, 0.50) | 49.95 [a] |
| Knowledge | 7 (3.31, 3.22, 0.32) | 26 (2.12, 2.18, 0.46) | 35.30 [a] |
| Work activities | 27 (3.60, 3.64, 0.39) | 14 (2.44, 2.44, 0.25) | 48.9 [a] |

[a] $p < 0.00001$.

**Table 2** Critically important skills, knowledge, and work activities for STEM disciplines (*High* group) listed from the largest (the top) to the smallest

| Skills (18) | Knowledge (7) | Work Activities (27) |
|---|---|---|
| <ul><li>Critical thinking</li><li>Reading comprehension</li><li>Active listening</li><li>Speaking</li><li>Complex problem solving</li><li>Judgment & Decision making</li><li>Writing</li><li>Monitoring</li><li>Active learning</li><li>Time management</li><li>Coordination</li><li>Systems analysis</li><li>Mathematics</li><li>Social perceptiveness</li><li>Systems evaluation</li><li>Instructing</li><li>Science</li><li>Learning strategies</li></ul> | <ul><li>English language</li><li>Mathematics</li><li>Computers & Electronics</li><li>Engineering & Technology</li><li>Administration & Management</li><li>Customer & personal service</li><li>Education & training</li></ul> | <ul><li>Getting Information</li><li>Making Decisions and Solving Problems</li><li>Interacting With Computers</li><li>Communicating with Supervisors, Peers, or Subordinates</li><li>Updating and Using Relevant Knowledge</li><li>Analyzing Data or Information</li><li>Identifying Objects, Actions, and Events</li><li>Processing Information</li><li>Documenting/Recording Information</li><li>Organizing, Planning, and Prioritizing Work</li><li>Thinking Creatively</li><li>Establishing and Maintaining Interpersonal Relationships</li><li>Evaluating Information to Determine Compliance with Standards</li><li>Interpreting the Meaning of Information for Others</li><li>Monitor Processes, Materials, or Surroundings</li><li>Communicating with Persons Outside Organization</li><li>Estimating the Quantifiable Characteristics of Products, Events, or Information</li><li>Judging the Qualities of Things, Services, or People</li><li>Training and Teaching Others</li><li>Scheduling Work and Activities</li><li>Developing Objectives and Strategies</li><li>Coordinating the Work and Activities of Others</li><li>Provide Consultation and Advice to Others</li><li>Developing and Building Teams</li><li>Inspecting Equipment, Structures, or Material</li><li>Coaching and Developing Others</li><li>Guiding, Directing, and Motivating Subordinates</li></ul> |

## A. Worker requirements in STEM disciplines

### *Important skills*

This section documents the skills that should be acquired or developed through experience and education for those following STEM careers. Of 35 skills, the importance of 18 skills was rated higher than 2.95 points for STEM disciplines (Table 3). Important skills in STEM disciplines are as follows (in order of importance; the number in parentheses represents the average importance): *Critical thinking (3.81), Reading comprehension (3.76), Active listening (3.75), Speaking (3.68), Complex problem solving (3.58), Judgment and decision making (3.51), Writing (3.51), Monitoring (3.40), Active learning (3.38), Time management (3.23), Coordination (3.19), System analysis (3.14), Mathematics (3.13), Social perceptiveness (3.12), Systems evaluation (3.07), Instructing (3.04), Science (2.97),* and *Learning strategies (2.96)* (for details, see Table A1).

To understand important skills, we reviewed operational definitions in the O*NET content model (see Table A1). First, STEM workers appear to be required to have higher-order thinking skills, such as *Critical Thinking, Complex Problem Solving,* and *Judgment and Decision Making.* Specifically, they are required to solve problems using skills of *Mathematics* and *Science*. STEM workers need to *use logic and reasoning to identify the strengths and weaknesses of alternative solutions*. Second, literacy skills, such as *Reading comprehension, Active Listening, Speaking,* and *Writing,* are required. STEM job incumbents need to *understand written sentences and paragraphs in work-related documents* and to *communicate effectively in writing as appropriate for the needs of the audience*. Third, they are required to solve problems relevant to organizations and systems with the skills underlying the descriptor *Monitoring, Systems Analysis, and Systems Evaluation.* They need to *identify measures or indicators of system performance and the actions needed to improve or correct performance, relative to the goals of the system*. Interestingly, competencies relevant to working with an organization or a system were not reported as important in frameworks of 21st century skills and engineering education as described in Tables 5 and 6. Fourth, STEM workers are required to frequently collaborate with others. With respect to interpersonal skills, skills such as *Coordination, Social perceptiveness,* and *Instructing* were reported as highly important. STEM job incumbents need to *adjust actions in relation to others' actions* and to *be aware of others' reactions and understand why they react as they do*. They are required to *teach others how to do something*. Fifth, competencies related to time management and updating of knowledge, that is *Time management and Learning Strategies,* are important. STEM job incumbents need to *manage* *one's own time and the time of others well* and *select and use training/instructional methods and procedures appropriate for the situation when learning or teaching new things*. To understand more about important competencies in terms of skills, see operational definitions in Table A1.

### *Important knowledge*

This section describes the knowledge that should be acquired or developed through experience and education for STEM careers. Table 2 shows the highly rated types of knowledge for STEM disciplines. Among 33 types of knowledge, the average importance of seven types of knowledge scored over 2.95 points. In order of importance, these were: *English language (3.74), Mathematics (3.70), Computers and Electronics (3.46), Engineering and technology (3.22), Administration and management (3.03), Customer and personal service (3.03),* and *Education and Training (2.98).* Among knowledge of natural sciences, *Physics (2.69)* was more important than *Chemistry (2.59)* and *Biology (2.37).* To understand important types of knowledge, please also see operational definitions in the O*NET in Table A2.

## B. Occupational requirements in STEM disciplines

### *Important work activities*

This section documents work activities that are considered of high importance for the STEM occupations. The highly rated work activities for STEM disciplines are listed in Table 2 (for details, see Table A3). The five most important work activities were: *Getting information (4.36), Making decisions and solving problems (4.18), Interacting with computers (4.14),* and *Communicating with supervisors, peers, or subordinates (4.08),* and *Updating and using relevant knowledge (4.04).* Most of the important work activities were relevant to dealing with and updating information and knowledge.

To understand important work activities, we reviewed operational definitions in the O*NET content model (see Table A3). First, STEM workers need high-level cognitive skills to get information and process information in order to solve problems. They solve problems by *breaking down information or data into separate parts* and *evaluating results to choose the best solution*. To think creatively, they need to *develop, design or create new applications, ideas, relationships, systems, or products, including artistic contributions*. They need to make decisions *whether events or processes comply with laws, regulations, or standards* with relevant information and experience. Additionally, STEM workers need competencies to use computers and equipment in order to *compile, code, categorize, calculate, verify information or data, write software, and set up functions*.

Second, STEM job incumbents need social communicative competencies ranging from *establishing and maintaining interpersonal relationships* to *guiding, leading, mentoring, training, and consulting other individuals*. In the workplace, they *develop and build a team with others to accomplish tasks*. To foster teamwork, they *encourage and build mutual trust, respect, and cooperate among team members*. They need to *identify the educational needs of others, develop formal educational or training programs or classes, and teach or instruct others*. They are required to provide *guidance and direction to subordinates, including setting performance standards and monitoring performance* and *advice to management or other groups on technical, systems-, or process-related topics*.

Third, STEM job incumbents need competencies relevant to working in an organizational system, such as *establishing long-range objectives and specifying the strategies and actions to achieve them* and *developing specific goals and planning to prioritize, organize, and accomplishing work*. These results show that competencies relevant to communication, interpersonal relationships, and leadership are needed, as well as competencies relevant to problem solving in Science, Technology, Engineering, and Mathematics. In the next section, we compare important competencies for STEM versus non-STEM disciplines.

## C. Relatively important skills, knowledge, and work activities

**Table 3** Skills, knowledge, and work activities of STEM disciplines with large effect sizes ($r > 0.3$) compared with non-STEM disciplines are ordered from the largest (top) to the smallest

| Effect size | Skills | Knowledge | Work Activities |
|---|---|---|---|
| $r > 0.3$ | <ul><li>Science</li><li>Mathematics</li><li>Programming</li><li>System Evaluation</li><li>System Analysis</li><li>Operations Analysis</li><li>Complex Problem Solving</li><li>Technology Design</li><li>Critical Thinking</li></ul> | <ul><li>Engineering & Technology</li><li>Mathematics</li><li>Physics</li><li>Computers and Electronics</li></ul> | <ul><li>Analyzing Data or Information</li><li>Estimating the Quantifiable Characteristics of Products, Events, or Information</li><li>Interacting With Computers</li><li>Processing Information</li></ul> |

Table 3 shows the relatively important skills, knowledge, and work activities identified for the STEM disciplines, with *r* scores over 0.3. Computing the effect size between two disciplines, we identified skills, knowledge, and work activities specifically important for STEM disciplines. For skills, nine descriptors rated much higher in STEM disciplines, as follows: *Science, Mathematics, Programming, Systems evaluation, Systems analysis, Operational analysis, Complex problem solving, Technology Design,* and *Critical thinking.* In knowledge, the following types were specifically regarded as being more important to STEM job incumbents as opposed to non-STEM counterparts: *Engineering and technology, Mathematics, Physics,* and *Computers and Electronics.* Relatively important work activities for STEM job incumbents were: *Analyzing data or information, Estimating the quantifiable characteristics of products, events, or information, Interpreting the meaning of information for others, and Processing information.* These results show STEM workers are required to have high-level skills for complex problem solving, information processing, and creative work with knowledge of science, mathematics, engineering, computing, and technology.

**D. Categorized important skills, knowledge, and work activities: STEM competencies**

Table 4 shows 52 important skills, knowledge, and work activities categorized into five domains using the framework developed by Katz and Kahn (1978) to illustrate general human performance. The inter-rater reliability (*Cohen's kappa*) between two raters was 0.74 ($p < 0.001$). After categorization, we reviewed operational definitions of skills, knowledge, and work activities of each domain in detail and renamed five to represent key competencies as follows: *(Ill-defined) Problem-solving skills, Social communication skills, Technology and engineering skills, System skills,* and *Time, resource, and knowledge management skills.*

First, descriptors relevant to problem solving, such as critical thinking, complex problem solving, knowledge of mathematics, skills of science, analyzing information, and thinking creatively, were categorized into *(Ill-defined) Problem solving skills.* Second, descriptors for communication in social contexts, such as speaking, coordination, knowledge of customer and personal service, and developing and building teams, were grouped into *Social communication skills.* These represent an essential component of performance in workplaces where most tasks are based on interpersonal communication. Third, *Technology and engineering skills*, such as programming, processing information, and practical application of engineering methods, are also essential in STEM workplaces.

**Table 4** Categorized important skills, knowledge, and work activities for the STEM disciplines. Eighteen skills, seven types of knowledge, and 27 work activities were grouped by two researchers into five categories of the framework developed by Katz and Kahn (1978): *Solving problems, working with people, working with technology, working with an organizational system,* and *working with resources.* The inter-rater reliability between two raters was 0.74

| Domain | Solving problems | Working with people | Working with technology | Working with an organizational system | Working with resources |
|---|---|---|---|---|---|
| 18 Skills | • Critical Thinking*<br>• Complex Problem Solving*<br>• Reading Comprehension<br>• Mathematics*<br>• Science* | • Active Listening<br>• Speaking<br>• Writing<br>• Coordination<br>• Social Perceptiveness<br>• Instructing | | • Monitoring<br>• Systems Analysis*<br>• Systems Evaluation*<br>• Judgment and Decision Making | • Time Management<br>• Learning Strategies<br>• Active Learning |
| 7 Knowledge | • Mathematics*<br>• English Language | • Customer and Personal Service<br>• Education and Training | • Computers & Electronics*<br>• Engineering & Technology* | • Administration and Management | |
| 27 Activities | • Getting Information<br>• Making Decisions & Solving Problems<br>• Analyzing Data or Information*<br>• Identifying Objects, Actions, and Events<br>• Thinking Creatively<br>• Evaluating Information to Determine Compliance with Standards | • Communicating with Supervisors, Peers, or Subordinates<br>• Establishing & Maintaining Interpersonal Relationships<br>• Interpreting the Meaning of Information for Others<br>• Communicating with Persons Outside Organization<br>• Training and Teaching Others<br>• Coordinating the Work & Activities of Others<br>• Provide Consultation & Advice to Others<br>• Developing and Building Teams<br>• Coaching and Developing Others<br>• Guiding, Directing, & Motivating Subordinates | • Interacting With Computers*<br>• Processing Information*<br>• Inspecting Equipment, Structures, or Material<br>• Documenting/Recording Information | • Monitor Processes, Materials, or Surroundings<br>• Judging the Qualities of Things, Services, or People<br>• Developing Objectives & Strategies | • Organizing, Planning, and Prioritizing Work<br>• Scheduling Work & Activities<br>• Updating and Using Relevant Knowledge<br>• Estimating the Quantifiable Characteristics of Products, Events, or Information* |
| STEM Competencies | (Ill-defined) Problem-solving skills | Social communication skills | Technology & engineering skills | System skills | Time, resource, and knowledge management skills |

\* Indicates relatively important skills, knowledge, and work activities in STEM disciplines with large effect sizes ($r > 0.3$) compared with non-STEM disciplines

**Table 5** Frameworks of twenty-first century skills in previous studies and deficiencies of domains compared with our results. Previous frameworks lack categories to explain a complete set of important skills, knowledge, and work activities for STEM occupations

| Framework | Author/impetus | Categories | Deficiency |
|---|---|---|---|
| 21st Century Skills | National Research Council (2008) | • Adaptability<br>• Complex Communication/Social Skills<br>• Non-routine Problem-solving Skills<br>• Self-management/Self-development<br>• Systems Thinking | Domains of working with technology based on knowledge of engineering and computing and working in and with an organization |
| The Assessment & Teaching of 21st Century Skills (ATC21S) | Collaboration among Cisco, Intel, Microsoft, the University of Melbourne, and others | • Ways of Thinking<br>• Ways of Working<br>• Tools for Working<br>• Living in the World | Domains of working with technology based on knowledge of engineering and computing and management time, resource, and knowledge |
| 21st Century Student Outcomes and Support systems (P21) | Partnership for 21st Century Skills (Several states and companies) | • Core Subjects and 21Century Skills<br>• Learning and Innovation Skills<br>• Information, Media, and Technology Skills<br>• Life and Career Skills | Domain of working in and with an organization |
| 21st-Century Competencies (revised) | Finegold and Notabartolo (2008) | • Analytic Skills<br>• Interpersonal Skills<br>• Ability to Execute<br>• Information Processing<br>• Capacity for Change | Domains of working with technology based on knowledge of engineering and computing, ill-defined problem solving, and working in and with an organization |

**Table 6** Categories of Engineering Criteria (2015-2016) have shortcomings when compared with important skills, knowledge, and work activities for STEM workplaces

| Framework | Abbreviated Title | Author/impetus | Categories | Deficiency |
|---|---|---|---|---|
| Engineering Criteria (2015-2016) | EC 2015-2016 | ABET | <ul><li>An ability to apply knowledge of science, math, engineering</li><li>An ability to design and conduct experiments, as well as to analyze and interpret data</li><li>An ability to design a system, component, or process to meet desired needs within realistic constraints such as economic, environmental, social, political, ethical, health and safety, manufacturability, and sustainability</li><li>An ability to function on multidisciplinary teams</li><li>An ability to identify, formulate, and solve engineering problems</li><li>An understanding of professional and ethical responsibility</li><li>An ability to communicate effectively</li><li>The broad education necessary to understand the impact of engineering solutions in a global and societal context</li><li>A recognition of the need for, and an ability to engage in lifelong learning</li><li>A knowledge of contemporary issues</li><li>An ability to use the techniques, skills, and modern engineering tools necessary for engineering practice</li></ul> | Domains of working with an organizational system, ill-defined problem solving, and time, resource, and knowledge management |

Fourth, with respect to working in and with an organization or a community (i.e., a sociotechnical system), a distinct set of *System skills* – monitoring processes, judging quality, and management – is needed to complete one's own tasks well. Fifth, *Time, resource, and knowledge management skills* may represent important influences on performance across a variety of job settings.

Five STEM competencies can be used to inform the skills and knowledge students should acquire through education and experience. STEM job incumbents must have knowledge that extends beyond science, technology, engineering, and mathematics. They need to solve ill-defined problems (using STEM knowledge), communicate with other professionals, understand how they work within an organization, and manage time, resources and knowledge. Of the five domains, the most common were *(Ill-defined) Problem-solving skills* and *Social communication skills*, which align with a result that shows cognitive skills and interpersonal skills are in demand in the current labor market (Autor et al. 2003). Findings from the current study reflect the required skills and knowledge in workplaces affected by rapid technological advances.

**E. Gaps between STEM competencies and present frameworks**

Comparing our results with frameworks of 21st century skills and engineering education, there are gaps between important STEM competencies and desirable outcomes of those frameworks. Table 5 lists categories of 21st century skills defined by four different frameworks, together with deficiencies compared with our findings. Frameworks of 21st century skills commonly lack categories relevant to *Technology and engineering skills*, *Time, resource, and knowledge management skills,* and *System skills* for working in an organizational system (Table 5). First, a framework of 21st Century Skills by the National Research Council (2008) suggests five necessary skill: adaptability, complex communication skills, non-routine problem-solving skills, self-management/development, and systems thinking (Koenig 2011). However, it did not illustrate all of important STEM competencies, such as *Technology and engineering skills* and *Time, resource, and knowledge management skills.* Second, the Assessment and Teaching of 21st Century Skills (ATC21S) organization developed the framework by synthesizing several national 21st century skills of the European Union, Organization for Economic Cooperation and Development (OECD), the United States and so on. ATC21S places skills into four categories: ways of thinking, ways of working, tools for working, and living in the world (Binkley et al. 2012). However, it does not contain categories to address *Technology and engineering skills* and *Time, resource, and knowledge management skills.* Third, P21 founded in 2002 with support from companies and the U.S. Department of Education, defined a framework for 21st century learning and suggests four important skill categories: core subjects and 21st century skills; learning and innovation skills; information, media, and technology skills; and, life and career skills (P21 2015). It does not include *System skills* relevant to working in the organizational system. Fourth, Finegold and Notabartolo (2010) developed a framework based on a review of the literature regarding required skills in future workplaces. They suggested five skills: analytic skills, interpersonal skills, ability to execute, information processing, and capacity for change (Finegold and Notabartolo 2010). They omitted important competencies relevant to working in the organization, ill-defined problem solving, and skills and knowledge of engineering and technology.

Further, the engineering criteria 2015-2016 (ABET 2015), which has determined the appropriate evaluation criteria for engineering education accreditation to better prepare students for the workplace, fails to include all important STEM competencies, such as *Ill-defined problem solving skills, System skills and Time, resource, and knowledge management skills* (Table 6). For example, engineering criteria suggest engineering graduates should have *"an ability to function on multidisciplinary teams",* yet it does not address the need for knowledge of administration and management. There is also no mention of time, resource, and knowledge management skills and ill-defined problem solving skills. Our findings align with studies that show the standards for engineering education are rarely successful in including required competencies from the perspective of practicing engineers, and might suggest a need to rethink the frameworks used for STEM education programs (Jonassen et al. 2006; Petroski 1996).

IV. **What is missing in 21st century skills and engineering criteria and what is needed?**

In this section, we discuss missing aspects of the frameworks through highlighting important STEM competencies. First, STEM careers require skills to solve ill-defined *problems* using knowledge of mathematics, science, and engineering. However, the engineering criteria (ABET 2015) are limited in their description of the ability to solve ill-structured problems as a learning outcome. It should be noted that workplace problems are typically unstructured and have multiple solution paths. For example, engineers need to deal with incomplete information and unanticipated problems, consider multiple sub-goals that often conflict with the primary goal, and use professional judgment to determine an optimal solution (Jonassen et al. 2006). Engineering criteria note that engineering graduates should have "*an ability to identify, formulate, and solve engineering problems*" (Table 6), yet they fail to address the ability to solve ill-defined problems (Felder and Brent 2003). Pólya (1945) identified steps in problem solving, suggesting that a problem-solver is required to deal with a whole process from getting information to examining the solution obtained. This finding suggests that a framework for STEM education, such as engineering criteria, is required to address the entire process of solving ill-defined problems from understanding an ill-defined problem to evaluating multiple solutions.

Second, STEM workers are required to be proficient in their knowledge of engineering, technology, and computing for data processing. *Technology and engineering skills* with knowledge of *Computers, Electronics, Engineering*, and *Technology* are important. Knowledge of engineering and technology are more important than knowledge of physics, chemistry, and biology in STEM workplaces. While frameworks relevant to 21st century skills have addressed key competencies needed for citizens and have influenced educational policy, curriculum innovation, and instructional practices, they have paid limited attention to STEM competencies. Frameworks for 21st century skills include *Information, Media, and Technology Skills* as required competencies for effective citizenship. However, they lack categories relevant to professional skills and knowledge of engineering, science, and technology. Indeed, all the important STEM competencies have yet to be detailed in one framework.

Third, knowledge and skills relevant to working within an organizational system, *System skills*, were found to be important. However, these skills were not noted as one of the desirable achievements in the engineering criteria (ABET 2015) and frameworks for 21st century skills. Interestingly, knowledge of *Administration and management* was more important to STEM employees than knowledge of physics, chemistry, and biology. Practicing engineers reported that most engineering problems require institutional knowledge about organizations, regulatory bodies, and support systems. This result suggests engineering criteria and a framework of STEM education are required to reflect the workplace context of most job incumbents in an organization.

Fourth, in the framework of 21st century skills, *Self- management/self-development* or *Life and career skills* are addressed as important. However, management of time, knowledge, and resources of self and others are not included. We identified *Time, knowledge, and resources management skills* as one of the important STEM competencies. Success of projects in STEM workplaces was rarely measured by engineering or technology standards alone, but typically included criteria related to time, budget, and satisfaction of customers (Jonassen et al. 2006). For efficient work process, they need to *estimate time, cost, resources, and materials needed to work*. With *scheduling work, activities, and time management of self and others*, STEM workers need to *organize, plan, and prioritize important work*. In addition to timing and budget, knowledge must also be managed. In a knowledge-based economy, the ability to acquire and manage knowledge is the hallmark of success (Smith 2001). Individuals without adequate knowledge, education, and training, struggle to keep up. A framework of 21st century skills and engineering criteria might need to address *time, knowledge, and resources management skills* to better prepare students for successful STEM careers.

Finally, we consider possible reasons for the gaps between our findings and the current frameworks. We assume there are gaps because of different approaches. Frameworks of 21st century skills and engineering criteria are developed by experts based on literature review or data collected from employers and educational leaders, rather than from data collected from employees. Perceptions of employers, educational leaders and employees about important skills, knowledge, and work activities may well differ. For example, STEM educational leaders might have difficulties recognizing that knowledge of management could be important for STEM careers. In this study, we analyzed O*NET, which informs skills, knowledge, abilities, and traits workers need based on job analysis. Our findings suggest educational leaders might need to consider making stronger links between educational curricula and required skills in the labor market, using job information to propose the best direction of STEM education.

## V.    Implications for STEM education

The President's Council of Advisors on Science and Technology (PCAST) has stressed, "STEM education… will determine whether the United States will remain a leader among nations and whether we will be able to solve immense challenges in areas such as energy, health, environmental protection, and national security" (p. 1) (PCAST 2010). For economic growth, the position of the leader of the world's STEM education is an important issue. Suppose that educational institutions do not supply an adequate number and quality of STEM-trained individuals for the workforce to meet social needs. Employers will hire them from other countries or move business offshore. Above all, from the perspective of individuals, education must serve to better prepare graduates for their careers. At the individual level, accumulating knowledge and knowhow is difficult because learning requires practice. However, STEM educators can better prepare students for successful STEM careers through well-designed educational programs.

In classrooms, students should be motivated to solve integrated, interdisciplinary sets of complex problems collaboratively using critical thinking and knowledge of STEM disciplines (Ahern-Rindell 1998; Hurd 1998). STEM education programs have successfully implemented problem-based learning (Felder and Brent 2003; Prados et al. 2005). More engineering education has been implemented as the National Assessment Governing Board has approved the evaluation of technology and engineering education through examinations given to US students from 2014 (Bybee 2010; NAEP 2013). However, problem solving that addresses ill-defined problems and demands the evaluation of multiple solution paths should be more encouraged in STEM education programs. The important role of education in promoting real-world problem solving skills has been addressed in studies of physics, mathematics, business, and engineering education (Ogilvie 2009; Adams and Wieman 2011; Fortus 2009; Robinson 2008; Shakerin 2006). STEM education programs might give more opportunities for students to solve ill-structured problems in classroom activities using qualitative (Facione 1990) and quantitative critical thinking (Holmes et al. 2015), and to participate in research projects relevant to real-world problems under supervision of professional scientists such as undergraduate research programs (Russell et al. 2007).

In the workplace, problem solving requires extensive collaboration because individuals cannot solve complex problems independently. Indeed, professionals rely on the collective knowledge of team members to solve problems and complete projects because of the size of the companies and problems (Jonassen et al. 2006). Collaborations can be successful when team members share a common goal and each member's role and relationship are well defined. In classrooms, collaborative problem solving can be implemented not only in learning activities but also in assessments. Exams that combine collaboration and testing, e.g., two-stage exams, can be used. A common approach is to add a collaborative component to the traditional individual exam so that students can discuss the individual questions immediately after completing the exam. A number of studies reported positive effects of collaborative exams: improved performance (Heller and Hollabaugh 1992; Clair and Chihara 2012; Gilley and Clarkston 2014), decreased test anxiety (Lusk and Conklin 2003), improved relationships with class mates (Shindler 2004), increase in retention (Stearns 1996), and increase in students' and instructors' positive perception (Stearns 1996; Wieman et al. 2014). Furthermore, well-designed group work has positively impacted student attitudes toward science, especially for female and ethnic minority students (Oakes 1990).

STEM education programs can promote communication skills as students practice technical writing and speaking in learning contexts [5, 52, 53]. STEM job incumbents are required to have professional skills in technical writing and speaking. Influenced by the ABET engineering criteria, the approach of writing-across-the-curriculum (WAC) has gained favor in engineering graduate education (Gunnink and Bernhardt 2002). However, gaps still exist between required communication skills and STEM education programs. Researchers have reported that students of science and technology in the upper grades of high school and in college have great difficulty reading scientific and mathematical texts [54]. Of the college students, 50% lacked the basic level of reading skill: the ability to read all the statements in a passage without skipping any [55]. Engineers recommended additional instruction on making oral presentations and writing, rather than more engineering in the engineering curriculum (Jonassen et al. 2006). Our study suggests that STEM education programs can create a continuous cycle where students practice communicating in learning contexts (Gray et al. 2005) and get frequent professional feedback from peers and educators using a peer- and self-assessment for writing, speaking and collaboration. This system for getting frequent feedback might promote students' learning substantially (Black and Wiliam 1998).

With respect to concerns about skills of the STEM workforce, students need to be encouraged to know the required competencies for their disciplines. It was reported that recruiters and students have significantly different perceptions regarding what constitutes important knowledge and skills for entry-level employees (Lee and Fang 2009). The greatest obstacles for college students in obtaining their first job included lack of job availability and knowledge about their own qualifications, skills, experiences, and personal qualities. Lack of information was also the biggest barrier for college students in their career development (Swanson and Tokar 1991). Students may be able to make better decisions about their career development when they know their own skills and important competencies required.

VI.    **Conclusion**

To better prepare students for the workplace, it is necessary for STEM educators to understand what STEM job incumbents do in their workplaces. Students must also recognize important competencies for their future careers. This study aimed to identify important 21st century STEM competencies using data from the workplace, and to consider these against the backdrop of current frameworks. Using a framework developed by Katz and Kahn (1978), 52 skills, types of knowledge, and work activities were categorized into five domains of important competencies: *(Ill-defined) Problem-solving skills, Social communication skills, Technology and engineering skills, System skills,* and *Time, resource, and knowledge management skills.* Results show that current frameworks do not comprehensively cover all the important STEM competencies, notably *problem solving skills (for ill-defined problems), System skills, Technology and engineering skills*, and *Time, resource, and knowledge management skills*. Our findings raise the possibility that present frameworks are inadequate in supporting STEM education programs to prepare students for their future careers and bridge gaps between education and required workplace skills. The role of education on subsequent career advancement has been addressed in the international setting (UNESCO 1996; OECD 2013). Educators enter the teaching profession to help young people learn, and their greatest rewards are when these goals, in this case helping students better prepare for future careers, are accomplished (Frase 1992). However, traditional curricula seem to reflect what teachers regard as important rather than what skills are actually required. As Hurd (1998) noted, the need to link education and work is essential for not only economic development but also the welfare of people and the quality of life (Hurd 1998). The framework of important STEM competencies (Table 6) presented here must still be considered as a tentative one. Given the empirical data and discussion, however, we are in a better position to support STEM education programs that focus on important competencies detailed by this framework so that ultimately students can be better prepared for their careers.

**Acknowledgement**

The author would like to thank Mazur group in Harvard University, especially Professor Eric Mazur for discussion, encouragement and support; Professor Hyewon Kim, Jung Bog Kim, and Minsu Ha for important, practical feedback on analysis; and physics education researchers and educators in the workshop of *Physics careers and majors* in AAPT 2015 summer meeting.

**Appendix A**
**Table A1** Important skills in STEM disciplines: operational definitions, mean, and standard deviations

| Important Skills | Operational definition | Mean (SD) |
|---|---|---|
| Critical Thinking | Using logic and reasoning to identify the strengths and weaknesses of alternative solutions, conclusions or approaches to problems | 3.81 (0.30) |
| Reading Comprehension | Understanding written sentences and paragraphs in work related documents | 3.76 (0.39) |
| Active Listening | Giving full attention to what other people are saying, taking time to understand the points being made, asking questions as appropriate, and not interrupting at inappropriate times | 3.75 (0.33) |
| Speaking | Giving full attention to what other people are saying, taking time to understand the points being made, asking questions as appropriate, and not interrupting at inappropriate times | 3.68 (0.38) |
| Complex Problem Solving | Identifying complex problems and reviewing related information to develop and evaluate options and implement solutions. | 3.58 (0.33) |
| Judgment & Decision Making | Considering the relative costs and benefits of potential actions to choose the most appropriate one. | 3.51 (0.32) |
| Writing | Communicating effectively in writing as appropriate for the needs of the audience | 3.51 (0.43) |
| Monitoring | Monitoring/Assessing performance of yourself, other individuals, or organizations to make improvements or take corrective action. | 3.40 (0.28) |
| Active Learning | Understanding the implications of new information for both current and future problem-solving and decision-making. | 3.38 (0.37) |
| Time Management | Managing one's own time and the time of others. | 3.23 (0.26) |
| Coordination | Adjusting actions in relation to others' actions. | 3.19 (0.29) |
| Systems Analysis | Determining how a system should work and how changes in conditions, operations, and the environment will affect outcomes. | 3.14 (0.45) |
| Mathematics | Using mathematics to solve problems. | 3.13 (0.59) |
| Social Perceptiveness | Being aware of others' reactions and understanding why they react as they do. | 3.12 (0.40) |
| Systems Evaluation | Identifying measures or indicators of system performance and the actions needed to improve or correct performance, relative to the goals of the system. | 3.07 (0.41) |
| Instructing | Teaching others how to do something. | 3.04 (0.48) |
| Science | Using scientific rules and methods to solve problems. | 2.97 (0.85) |
| Learning Strategies | Selecting and using training/instructional methods and procedures appropriate for the situation when learning or teaching new things. | 2.96 (0.47) |

**Table A2** Important knowledge in STEM disciplines: operational definitions, mean, and standard deviations.

| Knowledge | Operational Definition | Mean (SD) |
|---|---|---|
| English Language | Knowledge of the structure and content of the English language including the meaning and spelling of words, rules of composition, and grammar. | 3.74 (0.53) |
| Mathematics | Knowledge of arithmetic, algebra, geometry, calculus, statistics, and their applications. | 3.70 (0.64) |
| Computers & Electronics | Knowledge of circuit boards, processors, chips, electronic equipment, and computer hardware and software, including applications and programming. | 3.46 (0.69) |
| Engineering & Technology | Knowledge of the practical application of engineering science and technology. This includes applying principles, techniques, procedures, and equipment to the design and production of various goods and services. | 3.22 (1.09) |
| Administration & Management | Knowledge of business and management principles involved in strategic planning, resource allocation, human resources modeling, leadership technique, production methods, and coordination of people and resources. | 3.03 (0.51) |
| Customer & Personal Service | Knowledge of principles and processes for providing customer and personal services. This includes customer needs assessment, meeting quality standards for services, and evaluation of customer satisfaction. | 3.03 (0.67) |
| Education and Training | Knowledge of principles and methods for curriculum and training design, teaching and instruction for individuals and groups, and the measurement of training effects. | 2.98 (0.68) |

**Table A3** Important work activities in STEM disciplines: operational definitions, mean, and standard deviations.

| Important Work Activities | Operational Definitions | Mean (SD) |
|---|---|---|
| Getting Information | Observing, receiving, and otherwise obtaining information from all relevant sources | 4.36 (0.30) |
| Making Decisions and Solving Problems | Analyzing information and evaluating results to choose the best solution and solve problems | 4.18 (0.34) |
| Interacting With Computers | Using computers and computer systems (including hardware and software) to program, write software, set up functions, enter data, or process information. | 4.14 (0.63) |
| Communicating with Supervisors, Peers, or Subordinates | Providing information to supervisors, co-workers, and subordinates by telephone, in written form, e-mail, or in person | 4.08 (0.34) |
| Updating and Using Relevant Knowledge | Keeping up-to-date technically and applying new knowledge to your job. | 4.04 (0.43) |
| Analyzing Data or Information | Identifying the underlying principles, reasons, or facts of information by breaking down information or data into separate parts. | 3.95 (0.60) |
| Identifying Objects, Actions, and Events | Identifying information by categorizing, estimating, recognizing differences or similarities, and detecting changes in circumstances or events. | 3.95 (0.31) |
| Processing Information | Compiling, coding, categorizing, calculating, tabulating, auditing, or verifying information or data. | 3.91 (0.48) |
| Documenting/Recording Information | Entering, transcribing, recording, storing, or maintaining information in written or electronic/magnetic form. | 3.83 (0.47) |
| Organizing, Planning, and Prioritizing Work | Developing specific goals and plans to prioritize, organize, and accomplish your work. | 3.81 (0.32) |
| Thinking Creatively | Developing, designing, or creating new applications, ideas, relationships, systems, or products, including artistic contributions. | 3.71 (0.58) |
| Establishing and Maintaining Interpersonal Relationships | Developing constructive and cooperative working relationships with others, and maintaining them over time. | 3.71 (0.42) |
| Evaluating Information to Determine Compliance with Standards | Using relevant information and individual judgment to determine whether events or processes comply with laws, regulations, or standards. | 3.64 (0.57) |
| Interpreting the Meaning of Information for Others | Translating or explaining what information means and how it can be used. | 3.64 (0.57) |
| Monitor Processes, Materials, or Surroundings | Monitoring and reviewing information from materials, events, or the environment, to detect or assess problems. | 3.62 (0.50) |
| Communicating with Persons Outside Organization | Communicating with people outside the organization, representing the organization to customers, the public, government, and other external sources. This information can be exchanged in person, in writing, or by telephone or e-mail. | 3.46 (0. 65) |
| Estimating the Quantifiable Characteristics of Products, Events, or Information | Estimating sizes, distances, and quantities; or determining time, costs, resources, or materials needed to perform work activity. | 3.40 (0.49) |
| Judging the Qualities of Things, Services, or People | Assessing the value, importance, or quality of things or people. | 3.32 (0.40) |
| Training and Teaching Others | Identifying the educational needs of others, developing formal educational or training programs or classes, and teaching or instructing others. | 3.30 (0.60) |
| Scheduling Work and Activities | Scheduling events, programs, and activities, as well as the work of others. | 3.30 (0.47) |
| Developing Objectives and Strategies | Establishing long-range objectives and specifying the strategies and actions to achieve them. | 3.30 (0.49) |
| Coordinating the Work and Activities of Others | Getting members of a group to work together to accomplish tasks. | 3.21 (0.46) |

| Provide Consultation and Advice to Others | Providing guidance and expert advice to management or other groups on technical, systems-, or process-related topics. | 3.20 (0.57) |
|---|---|---|
| Developing and Building Teams | Encouraging and building mutual trust, respect, and cooperation among team members. | 3.13 (0.49) |
| Inspecting Equipment, Structures, or Material | Inspecting equipment, structures, or materials to identify the cause of errors or other problems or defects. | 3.07 (0.81) |
| Coaching and Developing Others | Identifying the developmental needs of others and coaching, mentoring, or otherwise helping others to improve their knowledge or skills. | 3.03 (0.53) |
| Guiding, Directing, and Motivating Subordinates | Providing guidance and direction to subordinates, including setting performance standards and monitoring performance. | 3.00 (0.54) |